\documentclass[onecolumn,prv]{revtex4}
\usepackage{graphicx}
\usepackage{amsmath}

\begin{document}
\title{Solvable Limit For SU(N) Kondo Model}
\author{Solomon F. Duki{\footnote{Currently at the University of Pittsburgh, Department of
Chemical Engineering, Pittsburgh PA, 15261}}}
\affiliation{Department of Physics, Case Western Reserve University, 
10900 Euclid Avenue, Cleveland, Ohio 44106-7079 \\}

\begin{abstract}
We study a single channel one dimensional Kondo Model where the impurity spin is 
replaced by an su(n) spin. Using Abelian bosonization and canonical transformation 
we explicitly show that this system has an exactly solvable point. The calculation 
also shows that there are $n$ collective excitation modes in the system, 
one charged and $n-1$ neutral spin excitation modes.  
\end{abstract}

\maketitle

\section{Introduction}

The Kondo problem \cite{Anderson:2005} and its subsequent multichannel generalization \cite{Nozieres:1980}
is a classic problem of condensed matter physics. 
Over the years different approaches have been used to address both the single and the
generalized multi-channel Kondo problem. This classic problem is now considered to be 
one of the class of condensed matter physics problems where a local degree of freedom 
interacts with a gap-less continuum. Some of the more powerful methods 
applied to understand properties of Kondo systems includes the renormalization group 
(RG) theory \cite{Nozieres:1980,Gan:1993,Lloyd:1981}, boundary conformal field theory (BCFT) 
\cite{Affleck:1991}, an exact solution by Bethe Ansatz \cite{Andrei:1980,Wiegmann:1980,
Andrei:1984}, exact solutions using bosonization and canonical transformations 
\cite{Toulouse:1970,Emery:1992,Fabrizio:1994}, and numerical methods \cite{Wilson:1975}. 

With the advancement of new methods in micro-fabrication and other 
experimental techniques enabled physicists to design and fabricate artificial 
atoms in nano-structures. These developments renewed the interest in Kondo physics in 
novel heterostructures, where the effect can be observed when an artificial magnetic 
impurity sits on an artificial metal (a two dimensional electron gas). Such experiments 
have been conducted using semiconductor quantum dots (SCQD), such as GaAs/AlGaAs and 
carbon nanotube quantum dots (CNQD) \cite{Gordon:1998,Jarillo:2005}. In these experiments, 
a tuneable magnetic impurity is formed by controlling the tunneling of electrons between 
the artificial atom and the 2D electron gas.

The conventional Kondo problem has a spin rotation or su(2) symmetry. However,
in nano-structures other higher symmetries are also possible, either due to additional 
internal degrees of freedom, or because of the way these hetero-structures are built 
in. In particular, there is growing interest in su(4) symmetry both in SCQD 
\cite{Borda:2003} and CNQD \cite{Choi:2005}, the case relevant to carbon nanotubes.

In this paper we study a single channel Kondo system that has an su(n) symmetry. 
It was discovered by Toulouse \cite{Toulouse:1970} that the conventional su(2) Kondo 
model has a simple solvable limit in the parameter space of the coupling constants. 
The su(2) Toulouse solution was subsequently extended to provide useful insights into 
the multichannel and Kondo lattice problems \cite{Emery:1992,Zachar:1996}. Here we 
demonstrate that the single channel Kondo model, with a generalized su(n) symmetry, has 
an exactly solvable limit. We begin by solving the su(4) model before turning our 
attention to the generalized su(n) model.

\section{The SU(n) Kondo Model}
\noindent We consider a single channel wire where electrons in the lead are assumed to
be non-interacting. The magnetic impurity is placed at the center of the wire so that 
it interacts with the free electron gas in the metal via exchange coupling. As we 
are interested in higher symmetries we assume electrons to have $n$ internal degrees 
of freedom. The case $n =2$ corresponds to an electron with spin. Higher $n$ values 
result if the electronic states are labeled by a sub-band index, as in the case of 
nanotubes where the orbital degeneracy is denoted by $+$ and $-$, or by a valley 
index, as in the case of silicon. We denote the Hamiltonian of the Fermi sea by 
$H_{0}$ and the exchange interaction of the impurity and the free electron gas by 
$H_{Kondo}$. Since the critical behavior of the Kondo system depends mainly on the 
interaction of the impurity and the $s$ angular momentum state of the Fermi 
sea, the radial equation can be used to describe the system. Following 
Schotte and Schotte \cite{Schotte:1969} we write the linearized  Hamiltonian in terms of chiral 
left moving fermions $\psi_{\alpha}(x)$ as\footnote{See Appendix A for the derivation.}
\begin{equation}
H=H_0+H_{Kondo},
\label {eq:H}
\end{equation}
where the kinetic energy is given by
\begin{equation}
H_{0}=\sum_{\alpha=1}^{n}\!\int_{-\infty}^{\infty}\psi_{\alpha}^{\dagger}(x)
(-i\partial_x)\psi_{\alpha}(x) dx
\label {eq:Hi_zero}
\end{equation}
and the exchange term  has the form  
\begin{equation}
H_{Kondo}=\sum_{\nu = 1}^{n^2 - 1} J_{\nu} S^{\nu} \tau^{\nu}.
\label{eq:H_int}
\end{equation}
Here we are working in units of $\hbar$=$v_{F}$=1, where $v_{F}$ is the Fermi velocity.
$\vec{\tau}$ is the su(n) impurity ``spin'' and 
\begin{equation}
\vec{S} = \sum_{\alpha, \beta=1}^{n} 
\psi_{\alpha}^{\dagger}(0)\vec{\Sigma}_{\alpha\beta}
\psi_{\beta}(0)
\label{eq:elecdensity}
\end{equation}
is the su(n) ``spin''
density of the conduction electrons at the origin. $J_{\nu}$ 
is the exchange coupling, which we assume to be independent of energy 
and the $\vec{\Sigma}$'s are the $n \times n$ traceless Hermitian 
matrices that represent the su(n) ``spin'' 
operators. These are a set of $n^2 - 1$ matrices that constitute the
basis for the set of $n \times n$ traceless hermitian matrices. 
Evidently $n-1$ of them are diagonal. They satisfy the ``orthogonality'' condition 
\begin{equation}
Tr(\Sigma_{\alpha}\Sigma_{\beta})=2\delta_{\alpha\beta}.
\label {eq:orthogonality}
\end{equation}
The $\Sigma$ matrices are called the {\em {Pauli}} matrices in su(2) case, 
the {\em {Gell-mann}} matrices for su(3), etc. 
\section{Toulouse Limit for SU(4) Model}
We now focus on the su(4) case to find its solvable limit. Later on we use similar 
formalism to generalize the result to the su(n) case. In su(4) symmetry the Hilbert 
space of the $4\times 4$ 
spin space can be spanned by the fifteen traceless $\Sigma$ matrices. We choose the 
three diagonal $\Sigma$ matrices which satisfy eq (\ref{eq:orthogonality}) as 
\begin{equation}
\begin{array}{c}
D_1 ={\frac {1}{\sqrt{2}}}
\left(
\begin{array}{cccc}
1 &   0 & 0 & 0 \\
0 &  -1 & 0 & 0 \\
0 &   0 & 0 & 0 \\
0 &   0 & 0 & 0 
\end{array} \right) \\
\\
D_2 ={\frac {1}{\sqrt{6}}}
\left( \begin{array}{cccc}
1 &  0 &  0 & 0 \\
0 &  1 &  0 & 0 \\
0 &  0 & -2 & 0 \\
0 &  0 &  0 & 0 
\end{array} \right) \\ 
\\
D_3 ={\frac {1}{\sqrt{12}}}
\left( \begin{array}{cccc}
1 &  0  & 0 & 0 \\
0 &  1  & 0 & 0 \\
0 &  0  & 1 & 0 \\
0 &  0  & 0 & -3 
\end{array}
\right) . 
\end{array}
\label{diagmatrix}
\end{equation}
The twelve off-diagonal matrices are selected from the matrices $O(\alpha,\beta)$ and $\tilde{O}
(\alpha,\beta)$
where  
\begin{equation}
\begin{array}{c}
O(\alpha,\beta)_{ij}=\delta_{\alpha{i}}\delta_{\beta{j}}+\delta_{\alpha{j}}\delta_{\beta{i}}\\
\tilde{O}(\alpha,\beta)_{ij}=-i(\delta_{\alpha{i}}\delta_{\beta{j}}-\delta_{\alpha{j}}
\delta_{\beta{i}}).
\end{array}
\label{offdiagonal_mat}
\end{equation}
These matrices are the generalizations of the {\em {Pauli}} matrices $\sigma_{x}$ and $\sigma_{y}$ 
and
we denote them by $O_{i}$ where $i=1,...12$ and  
\begin{equation}
\begin{array}{ccc}
O_1=O(1,2), & O_2=O(1,3), & O_3=O(1,4)\\
O_4=O(2,3), & O_5=O(2,4), &O_6=O(3,4)\\
O_7=\tilde{O}(1,2), & O_8=\tilde{O}(1,3), & O_9=\tilde{O}(1,4) \\ 
O_{10}=\tilde{O}(2,3), & O_{11}=\tilde{O}(2,4), & O_{12}=\tilde{O}(3,4).
\end{array}
\label{eq:offdiagonalelements}
\end{equation}
The $d$'s and the $O$'s, together, constitute the su(4) Lie Algebra. 
If the exchange coupling $J_{\nu}$ in eq (\ref{eq:H_int})
is independent of $\nu$ the Kondo model has a full su(n) symmetry.
Here we consider an anisotropic case for the exchange coupling where $J_{\nu}$ 
takes either $J_{\nu}=J_{\parallel}$ or $J_{\nu}=J_{\bot}$. This reduces the 
interaction part of the Hamiltonian into parallel and perpendicular components,  
\begin{equation}
H_{Kondo}=H_{Kondo}^{\parallel}+H_{Kondo}^{\bot}
\end{equation}
where
\begin{equation}
H_{Kondo}^{\parallel}=J_{\parallel}\sum_{\alpha,\beta=1}^{4}\sum_{\nu=1}^{3}\tau_{\parallel}^{\nu}
\psi^{\dagger}_{\alpha}(0)(D_{\nu})_{\alpha\beta}\psi_{\beta}(0)
\label{eq:parallel}
\end{equation}
and
\begin{equation}
H_{Kondo}^{\bot}=J_{\bot}\sum_{\alpha,\beta=1}^{4}\sum_{\nu=1}^{12}\tau_{\bot}^{\nu}
\psi^{\dagger}_{\alpha}(O_{\nu})_{\alpha\beta}\psi_{\beta}.
\label{eq:perpendicular}
\end{equation}
\section{Bosonization and Unitary Transformation}
The Hamiltonian of the system can take the form of a free Hamiltonian by 
Bosonizing the fermionic operators and then making a canonical transformation. 
Since the spin dynamics of the system depend only on the algebra that the spin 
operators satisfy, we prefer to work on the canonically transformed operators. The 
bosonization procedure can be done using the Mandelstam formula \cite{myron:1976,
Delft:1998,Shankar:1995}
where we can write chiral fermionic fields $\psi_{\alpha}$'s in terms of the bosonic fields 
$\phi_{\alpha}$'s as 
\begin{equation}
\psi_{\alpha}(x)={\frac{1}{\sqrt{2\pi{\epsilon}}}}e^{-i\phi^{-}_{\alpha}(x)},
\end{equation}
where
\begin{equation}
\phi^{-}_{\alpha}(x)={\sqrt{\pi}}\left[\int_{-\infty}^{x}dy\Pi_{\alpha}(y)+\phi_{\alpha}(x)\right].
\end{equation}
Here $\epsilon$ is the cutoff, which goes to zero in the continuum limit. $\Pi_{\alpha}(x)$ is the 
conjugate momentum of $\phi_{\alpha}(x)$ which satisfies the commutation relations 
\begin{equation}
\left[\phi_{\alpha}(x),\Pi_{\beta}(y)\right]=i\delta_{\alpha\beta}\delta(x-y).
\end{equation} 
For convenience we define the following excitations, which we call spin(s),
flavor(f), spin-flavor(fs) and charge(c) excitations as
\begin{eqnarray}
\begin{array}{rcl}
\phi^{-}_{s} & = &{\frac{1}{\sqrt{2}}}(\phi^{-}_{1} - \phi^{-}_{2}) \\
\phi^{-}_{f} & = &{\frac{1}{\sqrt{6}}}(\phi^{-}_{1} + \phi^{-}_{2} -2 \phi^{-}_{3}) \\
\phi^{-}_{sf} & = &{\frac{1}{\sqrt{12}}}(\phi^{-}_{1} + \phi^{-}_{2} + 
\phi^{-}_{3} - 3\phi^{-}_{4}) \\
\phi^{-}_{c} & = & {\frac{1}{2}}(\phi^{-}_{1} + \phi^{-}_{2} + \phi^{-}_{3} + \phi^{-}_{4}).
\end{array}
\label{eq:excitations}
\end{eqnarray}
Applying the bosonizing procedure in the free part of the Hamiltonian we have  
\begin{equation}
H_0={\frac{1}{2}}\sum_{\alpha=c,s,f,sf}\int_{-\infty}^{\infty}dx
\left[\left({\partial_x{\phi_{\alpha}^-(x)}}\right)^2+{{\Pi^-_{\alpha}}^2(x)}
\right].
\end{equation}
Similarly bosonization of the parallel part of the interaction Hamiltonian gives 
\begin{equation} 
H^{\parallel}_{Kondo}={\frac{J_{\parallel}}{\sqrt{\pi}}}\left.\left(\tau^{\parallel}_{1}(0)
{\frac{\partial{\phi^{-}_{s}}}{\partial{x}}}+\tau^{\parallel}_{2}(0){\frac{\partial
{\phi^{-}_{f}}}{\partial{x}}}+\tau^{\parallel}_{3}(0){\frac{\partial{\phi^{-}_{sf}}}
{\partial{x}}}\right)\right|_{x=0}.
\end{equation}
Bosonization of the perpendicular term, $H^{\bot}_{Kondo}$, leads to a more complicated
expression where spin-flip terms get coupled in pair wise fashion. However, these
$\tau^{\bot}$'s are coupled only through an effective rotations of 
$(\phi^{-}_{i}-\phi^{-}_{j})$, for $i\ne j$. Thus unitary transformation in the space 
of $\tau$ will remove the coupling. For a generic operator, $U$, its rotation is given by 
\begin{equation}
U(t)=e^{iFt}U(0)e^{-iFt}
\end{equation}  
where $t$ is a parameter and  $F$ is the generator of the unitary transformation. 
We choose this generator to be 
\begin{equation}
F=\left.\left(\tau^{\parallel}_{1}(0)\phi^{-}_{s}+
\tau^{\parallel}_{2}(0)\phi^{-}_{f}+
\tau^{\parallel}_{3}(0)\phi^{-}_{sf}\right)\right|_{x=0}.
\label{eq:generator}
\end{equation}  
Application of the canonical transformation on the bosonized $H^{\bot}_{Kondo}$ 
completely decouples the $\tau^{\bot}$'s at $t=\sqrt{4\pi}$; {\em {i. e.}}  
\begin{eqnarray}
H^{\bot}_{Kondo}(\sqrt{4\pi}) &  =  &
e^{iFt}~H^{\bot}_{Kondo}~e^{-iFt}\left|_{t=\sqrt{4\pi}}\right. \nonumber \\
& =  &  {\frac{J_{\bot}}{\pi{\epsilon}}}\sum_{\i=1}^{6}\tau^{\bot}_{2i-1}(0)
\label{eq:transformedH_perp}
\end{eqnarray}
The same canonical transformation on $H^{\parallel}_{Kondo}$ will give no 
additional terms. However, $H_0$  will be transformed in such a way that the
transformation of $H$ after bosonization can be written in the form
\footnote{For detailed derivations of eqns 
\ref{eq:transformedH_perp} and \ref{eq:decoupled} look at Appendix B.}   
\begin{eqnarray}
H & =  & e^{iFt}He^{-iFt}\left|_{t=\sqrt{4\pi}}\right. \nonumber \\ 
& = & {\frac{1}{2}}\sum_{k=c,s,f,sf}\int_{-\infty}^{\infty}dx
\left[\left({\partial_x{\phi_{k}^-(x)}}\right)^2+{{{\Pi}^-_{k}}^2(x)}\right] 
\nonumber\\
& & +\left({\frac{J_{\parallel}}{\sqrt{\pi}}}-t\right)\left.\left(\tau^{\parallel}_{1}(0)
{\frac{\partial{\phi^{-}_{s}}}{\partial{x}}}+\tau^{\parallel}_{2}(0){\frac{\partial
{\phi^{-}_{f}}}{\partial{x}}}+ \tau^{\parallel}_{3}(0){\frac{\partial{\phi^{-}_{sf}}}
{\partial{x}}}\right)\right|_{x=0} \nonumber\\
& & +{\frac{J_{\bot}}{\pi{\epsilon}}}\sum_{i=1}^{6}\tau^{\bot}_{2i-1}(0).
\label{eq:decoupled}
\end{eqnarray}
We clearly see that for $J_{\parallel}=2\pi$ the terms in the middle line of eq
(\ref{eq:decoupled}), which couples the free electron gas with the localized  
impurity spin, vanishes. Hence for $J_{\parallel}=2\pi$ the su(4) Kondo problem is
exactly solvable.
\section{Toulouse Limit for SU(n) Model}
A direct generalization of the same procedure reveals that the su(n) single channel 
Kondo model has the same solvable limit as that of the su(4) model, {\em {i.e.}} 
$J_{\parallel}=2\pi$. The su(n) generalization can be studied by bosonizing the 
Hamiltonian in eq (\ref{eq:H}) and extending eq (\ref{eq:generator}) to get the 
generalized form of the generator of the rotation in the $n$x$n$ dimensional 
matrix spin space. The appropriate choice for the generator is 
\begin{equation}
\mathcal{F}=\sum_{k=1}^{n-1}\tau^{\parallel}_{k}(0){\varphi}^{-}_{k}
\label{gen_generator}
\end{equation}
where the $\tau^{\parallel}_{k}$'s are the diagonal spin operators in their
representations and ${\varphi}^{-}_{k}$ are the $n$-1 different collective spin 
excitation modes, which are the generalizations of eq (\ref{eq:excitations}). Here we span
the spin space with $n^2-1$ hermitian matrices. As in the case of su(4) symmetry, a   
convenient choice of the $n-1$ diagonal matrices will be  
\begin{eqnarray}
\left[D_{k}\right]_{ij}=  {\frac{d_{k}^j}
{\displaystyle{\sqrt{{\sum_{j=1}^{n}}(d_{k}^j)^2}}}}\delta_{ij}
\end{eqnarray}
where 
\begin{equation}
d_{k}^j=\left\{
\begin{array}{ll}
1 & \textrm{if $j<k+1$}\\
-k & \textrm{if $j=k+1$}\\
0 & \textrm{if $j>k+1$}
\end{array}.
\right.
\end{equation}
The $d_{k}^j$'s are the $j^{th}$ elements of the $k^{th}$ diagonal 
matrix. The off-diagonal matrices are given by extending eq (\ref {offdiagonal_mat})
for the $n$x$n$ case. The collective spin excitation modes, ${\varphi}^{-}_{k}$, can 
be written in terms of the left moving Bose fields as
\begin{equation}
{\varphi}^{-}_{k} = \sum_{i=1}^{n} \left[D_{k} \right]_{ii} \phi_i^{-}
\end{equation}
and the charge mode is also given by
\begin{equation}
{\varphi}^{-}_{c}={\frac{1}{\sqrt{n}}}\sum_{k=1}^{n}\phi_{k}^{-}.
\end{equation}
The canonical transformation of the off-diagonal spin matrices $\tau^{\bot}$ is 
obtained from the evolution equation  
\begin{eqnarray}
-i{\frac{\partial}{\partial{t}}} {\tau^{\bot}_j(t)} =
e^{i\mathcal{F}t}[\mathcal{F},{\tau^{\bot}_j(0)}]e^{-i\mathcal{F}t}.
\label{eq:evolutionoftau_per}
\end{eqnarray}
Again here the spin operators and the $n^2-1$ hermitian matrices that span
the Hilbert space satisfy the same Lie Algebra, the commutator of ${\tau^{\bot}_j}$ 
and ${\tau^{\parallel}_k}$ can be obtained from the commutator of $D$'s and 
$O's({\tilde{O}}'s)$, which is given by  
\begin{eqnarray}
\left[O(j,k),D_{l}\right]=-i\left(d_{l}^{j}-d_{l}^{k}\right)\tilde{O}(j,k)\\ 
\left[\tilde{O}(j,k),D_{l}\right]=i\left(d_{l}^{j}-d_{l}^{k}\right)O(j,k)  
\label{eq:su(n)algebra}
\end{eqnarray}
where $O(j,k)$ and $\tilde{O}(j,k)$ are given by 
eq (\ref{offdiagonal_mat}).\\

Bosonization and canonical transformation of the su(n) Hamiltonian, eq (\ref{eq:H}), 
gives us 
\begin{eqnarray}
\lefteqn{
H={\frac{1}{2}}\sum_{k=1}^{n}\int_{-\infty}^{\infty}dx
\left[\left({\partial_x{\varphi_{k}^-(x)}}\right)^2+{{\Pi^-_{k}}^2(x)}\right] }\nonumber\\
& & +\left({\frac{J_{\parallel}}{\sqrt{\pi}}}-t\right)
\left.\sum_{k=1}^{n-1}\tau^{\parallel}_{k}
{\frac{\partial{\varphi^{-}_{k}}}{\partial{x}}}\right|_{x=0} \nonumber \\ 
& & +{\frac{J_{\bot}}{\pi{\epsilon}}}\sum_{i=1}^{{\frac{1}{2}}(n^2-n)}\left.
\tau^{\bot}_{2i-1}\right|_{t=0},
\label{solvable}
\end{eqnarray}
where again here we considered the energy independent anisotropic case of the exchange 
coupling, namely that  $J_{\nu}$ takes is either $J_{\nu}=J_{\parallel}$ or 
$J_{\nu}=J_{\bot}$. Clearly eq (\ref{solvable}) shows that for the model we considered 
the solvable point is the same as in the su(n) model.
\section{Summary and Conclusion}
In this work we have studied su(n) Kondo spin in a one dimensional single channel
wire with electrons in the lead assumed to be non-interacting. Using Abelian 
Bosonization of chiral fermions and canonical transformation we have found a 
solvable point for the problem, which is the su(n) generalization of the Toulouse 
limit \cite{Toulouse:1970}. This result may be used to test the large n approximation 
for the Kondo problem and a straightforward extension of this analysis can be applied 
to the multi-channel su(n) single impurity Kondo model and Kondo lattice problem.
Finally the exact solution obtained here may be used to compute the transport 
properties of nanostructures, a task to which we will return in future work.
\section*{Acknowledgment} 
I would like to thank Harsh Mathur for bring my attention to this problem and for his critical 
discussions through out this work. I would also like to thank Phil Taylor for valuable help.
\section*{Appendix A}
\section*{Linearization of Kondo Hamiltonian}
Here we show the linearization of the Hamiltonian of a single Kondo impurity in 
a formalism similar to that of Schotte and Schotte \cite{Schotte:1969}. 
Consider a field operator $\psi(x)$ which we write it in terms of its Fourier 
components as
\begin{eqnarray} 
\psi(x) & = &{\frac{1}{\sqrt{L}}}\sum_{k=-\infty}^{\infty}e^{ikx}c_k \nonumber \\
&=& {\frac{1}{\sqrt{L}}}\sum_{k=-\infty}^{-k_F}e^{ikx}c_k 
+ {\frac{1}{\sqrt{L}}}\sum_{k=-k_F}^{-k_F}e^{ikx}c_k 
+ {\frac{1}{\sqrt{L}}}\sum_{k=k_F}^{\infty}e^{ikx}c_k 
\label{eq:fieldoperator}
\end{eqnarray}
where $c^{\dagger}_k$ is electron creation operator and $k_F$ is the one dimensional
Fermi operator. Suppose we are interested only in the low lying excitations near
the Fermi surface. Then the field operator can be approximated as
\begin{eqnarray} 
\psi(x) & \approx &{\frac{1}{\sqrt{L}}}\sum_{p=-\Lambda}^{\Lambda}e^{i(k_F+p)x}c_{-(k_F+p)}
 + {\frac{1}{\sqrt{L}}}\sum_{p=-\Lambda}^{\Lambda}e^{i(k_F+p)x}c_{k_F+p}. 
\label{eq:fieldoperatorapproximated}
\end{eqnarray}
If we rename $c^{\dagger}_{k_F+p}=\alpha_{p}$ and $c^{\dagger}_{-(k_F+p)}=\beta_{p}$,
which creates electrons near the Fermi surface at $k=k_F$ and $k=k_{-F}$ respectively, 
then the approximated field operator in eq (\ref{eq:fieldoperatorapproximated}) 
can be written as 
\begin{eqnarray} 
\psi(x) = {\frac{1}{\sqrt{L}}}e^{ik_Fx}\psi_{+}(x) +
{\frac{1}{\sqrt{L}}}e^{-ik_Fx}\psi_{-}(x). 
\label{eq:linearizedfieldoperator}
\end{eqnarray}
where
\begin{eqnarray} 
\psi_{+}(x) = {\frac{1}{\sqrt{L}}}\sum_{p=-\Lambda}^{\Lambda}e^{ipx}\alpha_p 
\hspace{2mm}\textrm {and} \hspace{2mm}
\psi_{-}(x) = {\frac{1}{\sqrt{L}}}\sum_{p=-\Lambda}^{\Lambda}e^{-ipx}\beta_p.
\label{eq:linearizedfieldoperator1}
\end{eqnarray}
Here $L$ is the length of the 1D wire. The operators $\psi^{\dagger}_{+}$ and 
$\psi^{\dagger}_{-}$ are called right and left moving chiral fermionic operators 
respectively, for a reason that will become clear, from the Hamiltonian form, shortly. 
The Hamiltonian of a free electron gas can be written in terms of the left and 
right moving chiral fermions as
\begin{eqnarray} 
H_0 & = &\int dx ~ \psi^{\dagger}(x)\left(-{\frac{1}{2}}{\frac{\partial^2}{\partial x^2}}
\right)\psi(x)\nonumber \\ & = &  
k_F\int dx \left[ \psi_{+}^{\dagger}(x)(- i \partial_x)\psi_{+}(x) +
 \psi_{-}^{\dagger}(x)(i \partial_x)\psi_{-}(x) \right ] + \nonumber \\
& {} &  {\frac{k^2_F}{2}} + \textrm{highly oscillating terms}.
\label{eq:linearizedhamiltonian}
\end{eqnarray}
If we rescaled the energy with respect to the Fermi level and neglect highly 
oscillating terms we obtain a Hamiltonian whose form is similar to that of  
left- and right-handed massless fermions. In terms of $\alpha_{p}$ and  $\beta_{p}$
the Hamiltonian is given by
\begin{equation}
H_0 = {\frac{k^2_F}{2}} + k_F\sum_{p}\left[\alpha^{\dagger}_{p}\alpha_{p} -
\beta^{\dagger}_{p}\beta_{p} \right].
\end{equation}
\section*{Appendix B} 
\section*{Details of Canonical Transformation}
In this section we show how we derived eq (\ref{eq:transformedH_perp}) and 
(\ref{eq:decoupled}). We begin first with the construction of the su(4) spin
representations using Schwinger's method of oscillators. We assume that 
$d_{j}^{\dagger}$ creates an electron on the impurity site with spin state $|j\rangle$.
Then the su(4) spin space can be generated using the sixteen number 
conserving bilinear, $S_{ij}=d^{\dagger}_id_j$. For a base state $|\mu\rangle$ 
the bilinear acts according to $S_{ij}|\mu\rangle=\delta_{j\mu}| i \rangle$. 
The commutator of these bilinear is given by
\begin{equation}
[S_{ij},S_{kl}] = S_{il}\delta_{jk}-S_{jk}\delta_{il}.
\label{eq:su4algebra}
\end{equation}
Using these bilinear we write the number and spin operators as  
\begin{eqnarray}
\begin{array}{ll}
N=\displaystyle{\sum_{j}d^{\dagger}_{j}d_{j}} \\
\displaystyle{\tau^{\parallel}_{\mu}=\sum_{ij}d^{\dagger}_i[D_{\mu}]_{ij}d_j}
& \textrm{ for $\nu$ = 1,2,3 }\\
\displaystyle{\tau^{\bot}_{\nu}=\sum_{ij}d^{\dagger}_{i}[O_{\nu}]_{ij}d_j} 
& \textrm{ for $\nu$ = 1, $\ldots$ 12 }
\end{array}
\label{eq:su4spins}
\end{eqnarray}
where $O_{\nu}$ is given by eq (\ref{eq:offdiagonalelements}). Using the su(4)
algebra, eq (\ref{eq:su4algebra}), one can get the commutation 
$[\tau^{\parallel}_{\mu},\tau^{\bot}_{\nu}]$ for any $\mu$ and $\nu$. In fact,
these spin operators will satisfy the same algebra as the fifteen matrices
that we used to span the spin space in eqns (\ref{diagmatrix}) and 
(\ref{eq:offdiagonalelements}). 
We now consider the derivation of eq (\ref{eq:transformedH_perp}). Bosonization of 
$H^{\bot}_{Kondo}$ from eq (\ref{eq:perpendicular}) results in
\begin{eqnarray}
{\frac{\pi\epsilon}{J_{\bot}}}H^{\bot}_{Kondo} 
& = & 
\tau_1^{\bot}(0)\cos[\sqrt{4\pi}(\phi^{-}_{1}-\phi^{-}_{2})] + 
\tau_2^{\bot}(0)\sin[\sqrt{4\pi}(\phi^{-}_{1}-\phi^{-}_{2})] + \nonumber\\
& {} &
\tau_3^{\bot}(0)\cos[\sqrt{4\pi}(\phi^{-}_{1}-\phi^{-}_{3})] + 
\tau_4^{\bot}(0)\sin[\sqrt{4\pi}(\phi^{-}_{1}-\phi^{-}_{3})] + \nonumber\\
& {} &
\tau_5^{\bot}(0)\cos[\sqrt{4\pi}(\phi^{-}_{1}-\phi^{-}_{4})] +
\tau_6^{\bot}(0)\sin[\sqrt{4\pi}(\phi^{-}_{1}-\phi^{-}_{4})] + \nonumber\\
& {} &
\tau_7^{\bot}(0)\cos[\sqrt{4\pi}(\phi^{-}_{2}-\phi^{-}_{3})] - 
\tau_8^{\bot}(0)\sin[\sqrt{4\pi}(\phi^{-}_{2}-\phi^{-}_{3})] + \nonumber\\
& {} &
\tau_9^{\bot}(0)\cos[\sqrt{4\pi}(\phi^{-}_{2}-\phi^{-}_{4})] - 
\tau_{10}^{\bot}(0)\sin[\sqrt{4\pi}(\phi^{-}_{2}-\phi^{-}_{4})] + \nonumber\\
& {} &
\tau_{11}^{\bot}(0)\cos[\sqrt{4\pi}(\phi^{-}_{3}-\phi^{-}_{4})] -  
\tau_{12}^{\bot}(0)\sin[\sqrt{4\pi}(\phi^{-}_{3}-\phi^{-}_{4})].
\label{eq:bosonization_H_per}
\end{eqnarray}
A straight forward calculation of the commutation 
$[\tau^{\parallel}_{\mu},\tau^{\bot}_{\nu}]$ determines the evolution of the spin 
operators through eq (\ref{eq:evolutionoftau_per}); i.e.,
\begin{eqnarray}
-i{\frac{\partial}{\partial{t}}}{\tau_j^{\bot}}(t)=e^{iFt}[F,\tau_j^{\bot}(0)]e^{-iFt}
\nonumber
\end{eqnarray}
where $F$ is the generator of the rotation which was defined in eq (\ref{eq:generator}).
These differential equations are coupled in a pair wise fashion and their solution are given as
follows:
\begin{eqnarray}
\begin{array}{rcl}
\tau_1^{\bot}(t) & = & \tau_1^{\bot}(0)\cos[(\phi^{-}_{1}-\phi^{-}_{2})t] -
			\tau_2^{\bot}(0)\sin[(\phi^{-}_{1}-\phi^{-}_{2})t] \\
\tau_2^{\bot}(t) & = & \tau_2^{\bot}(0)\cos[(\phi^{-}_{1}-\phi^{-}_{2})t] +
			\tau_1^{\bot}(0)\sin[(\phi^{-}_{1}-\phi^{-}_{2})t] \\
\tau_3^{\bot}(t) & = & \tau_3^{\bot}(0)\cos[(\phi^{-}_{1}-\phi^{-}_{3})t] -
			\tau_4^{\bot}(0)\sin[(\phi^{-}_{1}-\phi^{-}_{3})t] \\
\tau_4^{\bot}(t) & = & \tau_4^{\bot}(0)\cos[(\phi^{-}_{1}-\phi^{-}_{3})t] +
			\tau_3^{\bot}(0)\sin[(\phi^{-}_{1}-\phi^{-}_{3})t] \\
\tau_5^{\bot}(t) & = & \tau_5^{\bot}(0)\cos[(\phi^{-}_{1}-\phi^{-}_{4})t] -
			\tau_6^{\bot}(0)\sin[(\phi^{-}_{1}-\phi^{-}_{4})t] \\
\tau_6^{\bot}(t) & = & \tau_6^{\bot}(0)\cos[(\phi^{-}_{1}-\phi^{-}_{4})t] + 
			\tau_5^{\bot}(0)\sin[(\phi^{-}_{1}-\phi^{-}_{4})t] \\
\tau_7^{\bot}(t) & = & \tau_7^{\bot}(0)\cos[(\phi^{-}_{2}-\phi^{-}_{3})t] +
			\tau_8^{\bot}(0)\sin[(\phi^{-}_{2}-\phi^{-}_{3})t] \\
\tau_8^{\bot}(t) & = & \tau_8^{\bot}(0)\cos[(\phi^{-}_{2}-\phi^{-}_{3})t] -
			\tau_7^{\bot}(0)\sin[(\phi^{-}_{2}-\phi^{-}_{3})t] \\
\tau_9^{\bot}(t) & = & \tau_9^{\bot}(0)\cos[(\phi^{-}_{2}-\phi^{-}_{4})t] +
			\tau_{10}^{\bot}(0)\sin[(\phi^{-}_{2}-\phi^{-}_{4})t] \\
\tau_{10}^{\bot}(t) & = & \tau_{10}^{\bot}(0)\cos[(\phi^{-}_{2}-\phi^{-}_{4})t] -
			\tau_9^{\bot}(0)\sin[(\phi^{-}_{2}-\phi^{-}_{4})t] \\
\tau_{11}^{\bot}(t) & = & \tau_{11}^{\bot}(0)\cos[(\phi^{-}_{3}-\phi^{-}_{4})t] +
			\tau_{12}^{\bot}(0)\sin[(\phi^{-}_{3}-\phi^{-}_{4})t] \\
\tau_{12}^{\bot}(t) & = & \tau_{11}^{\bot}(0)\cos[(\phi^{-}_{3}-\phi^{-}_{4})t] -
			\tau_{12}^{\bot}(0)\sin[(\phi^{-}_{3}-\phi^{-}_{4})t] 
\end{array}
\label{eq:spinevolution}
\end{eqnarray}
Application of the transformation $e^{iFt}~H^{\bot}_{Kondo}~e^{-iFt}$, which utilizes
eq (\ref{eq:spinevolution}), completely decouples the diagonal
and off-diagonal spin operators at $t=\sqrt{4\pi}$, giving the final result shown in
eq (\ref{eq:transformedH_perp}).

To make a canonical transformation on  $H_0$ and $H^{\parallel}_{Kondo}$ we first write
both of these terms in terms of the spin excitation fields. The time evolution of these 
field operators ($\varphi^{-}_{c},~\varphi^{-}_{s}, ~\varphi^{-}_{f}$ and $\varphi^{-}_{sf}$)
and their conjugate fields ($\Pi^{-}_{c},~\Pi^{-}_{s},~\Pi^{-}_{f}$ and $\Pi^{-}_{sf}$) are
determined by 
\begin{eqnarray}
\begin{array}{rcl}
-i{\frac{\partial}{\partial{t}}}{\varphi_j^{-}}(x,t) & = & e^{iFt}[F,\varphi_j^{-}(x)]e^{-iFt}\\
-i{\frac{\partial}{\partial{t}}}{\Pi_j^{-}}(x,t) & = & e^{iFt}[F,\Pi_j^{-}(x)]e^{-iFt}.
\end{array}
\label{filedevolution}
\end{eqnarray}
However, the commutation relation between the generator $F$ and the fields are give by
\begin{equation}
[F,\varphi_j^-(x)]  =  -i\tau^{\parallel}_k(0)\Theta(-x)\delta_{jk} 
\end{equation}
and
\begin{equation}
[F,\Pi_j^-(x)]   =  i\tau^{\parallel}_k (0) \delta (x)\delta_{jk}
\end{equation}
where  $\Theta(-x)$ is the Heaviside step function. Utilizing the solutions of 
eq (\ref{filedevolution}), one can show that the parallel component of the Kondo Hamiltonian 
is canonically transformed in to
\begin{eqnarray}
H^{\parallel}_{Kondo} & = & e^{iFt}H^{\parallel}_{Kondo}e^{-iFt}i \nonumber \\ 
&= &
\left.{\frac{J_{\parallel}}{\sqrt{\pi}}}\left(
\tau^{\parallel}_{1}(0){\frac{\partial{\phi^{-}_{s}}}{\partial{x}}}+
\tau^{\parallel}_{2}(0){\frac{\partial{\phi^{-}_{f}}}{\partial{x}}} + 
\tau^{\parallel}_{3}(0){\frac{\partial{\phi^{-}_{sf}}}{\partial{x}}}
\right)\right|_{x=0} \nonumber \\
& {} & +  \textrm{ diverging constant }.
\end{eqnarray}
Similarly the kinetic energy part is also transformed as  
\begin{eqnarray}
H_0 & = & e^{iFt}H_0e^{-iFt} \nonumber \\
& =  &
{\frac{1}{2}}\sum_{k=c,s,f,sf}\int_{-\infty}^{\infty}dx 
\left[\left({\partial_x{\phi_{k}^-(x)}}\right)^2+{{\Pi^-_{k}}^2(x)} \right] 
\nonumber \\
& {} & t\left.\left(
\tau^{\parallel}_{1}(0){\frac{\partial{\phi^{-}_{s}}}{\partial{x}}}+
\tau^{\parallel}_{2}(0){\frac{\partial{\phi^{-}_{f}}}{\partial{x}}} + 
\tau^{\parallel}_{3}(0){\frac{\partial{\phi^{-}_{sf}}}{\partial{x}}}
\right)\right|_{x=0} \nonumber \\
& {} & +  \textrm { diverging constant }
\label{eq:transformationofH_0}
\end{eqnarray}
We throw away the the diverging constant as it is a term that can be renormalized, and 
hence arrive at eq (\ref{eq:decoupled}). A straight forward, and similar, procedure is applied 
to get the solvable point of the su(n) Kondo Model, where the same algebra of 
eq (\ref{eq:su(n)algebra}) is used to get the commutator of the su(n) spin operators.

\end{document}